\documentclass[lettersize,journal]{IEEEtran}
\usepackage{amsmath,amsfonts,amssymb}
\usepackage{algorithmic}
\usepackage{algorithm}
\usepackage{array}
\usepackage[caption=false,font=normalsize,labelfont=sf,textfont=sf]{subfig}
\usepackage{textcomp}
\usepackage{stfloats}
\usepackage{url}
\usepackage{hyperref}
\usepackage{verbatim}
\usepackage{graphicx}
\usepackage{cite}
\usepackage{soul,xcolor}

\newcommand\aap{A\&A}%
\newcommand\apj{\ref@jnl{ApJ}}%
\newcommand\apjl{\ref@jnl{ApJ}}%

\newcommand\jcap{Journal of Cosmology and Astroparticle Physics}
\newcommand\ao{Applied Optics}

\newcommand{\nepunits}{W\,Hz$^{-1/2}$}

\newcommand{\gt}{>}

\newcommand{\micron}{$\mu$m}
\def\gsim{\mathrel{\rlap{\lower4pt\hbox{\hskip1pt$\sim$}}
    \raise1pt\hbox{$>$}}}
\def\lsim{\mathrel{\rlap{\lower4pt\hbox{\hskip1pt$\sim$}}
    \raise1pt\hbox{$<$}}}

\newcommand{\singlecolwidth}{3.25in}

\setstcolor{red}

\begin{document}

\title{Characterization of a Far-Infrared Kinetic Inductance Detector Prototype for PRIMA}

\author{Steven Hailey-Dunsheath, Sven van Berkel, Andrew D. Beyer, Logan Foote, Reinier M. J. Janssen, Henry G. LeDuc, Pierre M. Echternach, Charles M. Bradford, Jochem J. A. Baselmans, Shahab Dabironezare, Peter K. Day, Nicholas F. Cothard, Jason Glenn
\thanks{Funding for this work was provided by the National Aeronautics and Space Administration (Grant No. 80NSSC19K0489; PI S. Hailey-Dunsheath). This work was performed in part at the Jet Propulsion Laboratory, California Institute of Technology, under a contract with the National Aeronautics and Space Administration. Copyright 2023 California Institute of Technology. U.S. Government sponsorship acknowledged.}
\thanks{Steven Hailey-Dunsheath and Logan Foote are with the California Institute of Technology, Pasadena, CA 91125, USA (e-mail: haileyds@caltech.edu).}
\thanks{Sven van Berkel, Andrew D. Beyer, Reinier M. J. Janssen, Henry G. LeDuc, Pierre M. Echternach, Charles M. Bradford, and Peter K. Day are with the Jet Propulsion Laboratory, California Institute of Technology, Pasadena, CA 91109, USA.}
\thanks{Jochem J. A. Baselmans and Shahab Dabironezare are with the Faculty of Electrical Engineering, Mathematics and Computer Science, Delft University of Technology,
Mekelweg 4, 2628 CD Delft, The Netherlands, and with SRON—Netherlands Institute for Space Research, Niels Bohrweg 4, 2333 CA Leiden, The Netherlands.}
\thanks{Nicholas F. Cothard and Jason Glenn are with NASA Goddard Space Flight Center, 8800 Greenbelt Rd, Greenbelt, 20771, MD, USA.}}



\maketitle


\begin{abstract}
The PRobe far-Infrared Mission for Astrophysics (PRIMA) is under study as a potential far-IR space mission, featuring actively cooled optics, and both imaging and spectroscopic instrumentation. To fully take advantage of the low background afforded by a cold telescope, spectroscopy with PRIMA requires detectors with a noise equivalent power (NEP) better than $1 \times 10^{-19}$ \nepunits. To meet this goal we are developing large format arrays of kinetic inductance detectors (KIDs) to work across the $25-250$ {\micron} range. Here we present the design and characterization of a single pixel prototype detector optimized for $210$ {\micron}. The KID consists of a lens-coupled aluminum inductor-absorber connected to a niobium interdigitated capacitor to form a 2 GHz resonator. We have fabricated a small array with 28 KIDs, and we measure the performance of one of these detectors with an optical loading in the $0.01 - 300$ aW range. At low loading the detector achieves an NEP of $9\times10^{-20}$ \nepunits\ at a 10 Hz readout frequency. An extrapolation of these measurements suggests this detector may remain photon noise limited at up to 20 fW of loading, offering a high dynamic range for PRIMA observations of bright astronomical sources.
\end{abstract}





\section{Introduction} \label{sect:intro}

Far-infrared (far-IR; $25-250$ {\micron}) spectroscopy and spectrophotometry are unique tools for studying the interiors of galaxies and forming planetary systems. Far-IR radiation readily escapes regions that are deeply obscured by dust at optical and near-infrared wavelengths, and far-IR measurements can provide quantitative metrics of star formation and black hole growth in distant galaxies, as well as the chemical constituents of protoplanetary disks as they evolve.  Revolutionary advances are still possible with a modest space mission in the far-IR, because a platform that operates at the fundamental limits in the band between JWST and ALMA has not yet been fielded. We are developing the PRobe far-Infrared Mission for Astrophysics (PRIMA\cite{prima_go_book})\footnote{\url{www.prima.ipac.caltech.edu}}, a NASA concept that features a 2-m class telescope and optics cooled to below 5~K. With sufficiently sensitive detectors, this cold telescope enables measurements limited by only the photon shot noise of the zodiacal and Galactic dust emission. The PRIMA FIRESS grating spectrometer has an $R \sim 130$ spectral bin per detector pixel, and an estimated 20\% optical efficiency into a 1 f$\lambda$ dual-polarization pixel. When looking toward the north ecliptic pole, this fundamental photon noise limit corresponds to a detector noise equivalent power (NEP) of $\approx$$1\times$10$^{-19}$ \nepunits\ in the $100-250$ $\mu$m range.

Kinetic inductance detectors (KIDs) are poised to meet this demanding sensitivity requirement while also offering favorable system properties compatible with a modest space mission: operation temperature up to 150~mK, high multiplex factor, and low focal plane power dissipation. KIDs have demonstrated high yield in kilopixel arrays \cite{Baselmans2017,Liu_2022TIM}, sufficient immunity to cosmic-ray interactions \cite{Karatsu2019CR,kane_ltd2023}, and have been fielded successfully in balloon instruments \cite{Masi2019olimpo,Lowe_2020}. The unique sensitivity requirements for a PRIMA-like mission have been demonstrated in antenna-coupled 1.5~THz (200~\micron) devices \cite{baselmans2012ultra,Baselmans2022}, setting the stage for proceeding with PRIMA for implementation this decade.  Here we present a KID design developed specifically for PRIMA that uses a lens-coupled all aluminum multi-mode absorber architecture, building on the development of KID focal planes for the TIM and BLAST balloon experiments\cite{Liu_2022TIM,haileydunsheath2018,Janssen_2023,Lowe_2020}.  Relative to an antenna-coupled (single-mode) approach, the multi-mode absorber coupling can scale to the shortest far-IR wavelengths, and offers increased versatility in the instrument design. This article focuses on a 210 \micron\ KID that demonstrates an NEP below 10$^{-19}\,\rm W\,Hz^{-1/2}$ at 10 Hz.  We present the device design, measured performance, and validation of our KID sensitivity model, which allows the confident prediction of in-flight performance with PRIMA.  We have also designed and built 1008-pixel arrays using this pixel architecture. Early testing of these shows $>$90\% resonator yield; sensitivity statistics and cosmic-ray response are presented in companion articles \cite{foote_ltd2023,kane_ltd2023}.

\section{Detector Design and Fabrication} \label{sec:detector_design}

\subsection{Device Description}

\begin{figure}
\centering
\includegraphics[width=\singlecolwidth]{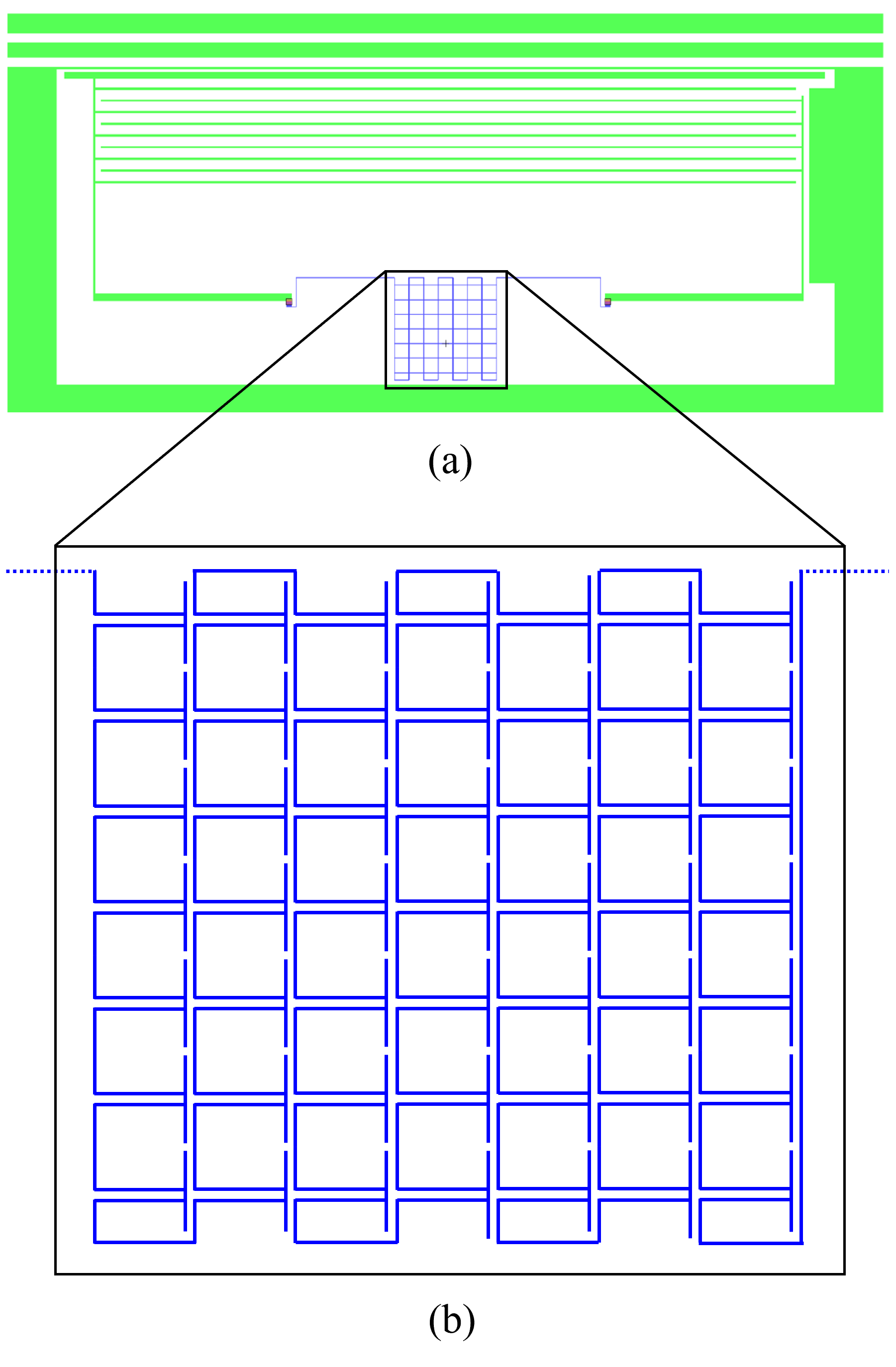} 
\caption{\small Layout of a single KID. (\textit{a}) Niobium features include the CPW readout line along the top of the figure, the IDC, and the surrounding ground plane (\textit{green}). The inductor-absorber is aluminum (\textit{blue}), and connects to the IDC through a niobium plug (\textit{red}). (\textit{b}) Schematic of the inductor-absorber highlighting the connections between individual unit cells. This drawing is not to scale -- the gaps between traces have been exaggerated for clarity (see Figure \ref{fig:absorber_fig}).
\label{fig:kid_schematic}
}
\end{figure}

The layout of an individual KID is shown in Figure \ref{fig:kid_schematic}. Each pixel consists of a niobium interdigitated capacitor (IDC) connected to an aluminum inductor to form a parallel LC resonant circuit. The IDC uses 2~\micron\ wide fingers with a 10~\micron\ spacing, and the capacitor size is varied from pixel to pixel so each pixel will have a unique resonance frequency, allowing for a frequency multiplexed readout\cite{foote_ltd2023}. A niobium coplanar waveguide (CPW) structure meanders through the device to capacitively couple the microwave readout signal to each individual pixel. A region of the ground plane extends to the edge of the IDC to provide a capacitive coupling, and a path to ground for the microwave current. The feedline and IDC are made from a sputter-deposited layer of niobium, and patterned using reactive ion etching. During etching, a resist mask patterned using optical stepper lithography is used to achieve the desired geometry. The aluminum inductor is defined by electron beam lithography and liftoff of a 30~nm thick sputtered aluminum layer. The connection between the niobium IDC and the aluminum inductor is accomplished by a niobium plug defined by photolithography. Prior to the niobium deposition the surface layer of aluminum oxide is stripped using ion milling to provide a good electrical contact between the capacitor and inductor. The device is fabricated on a 300 {\micron} thick high resistivity float zone (HRFZ) grown silicon wafer to provide good transmission in the far-IR. This wafer is then glued to another HRFZ silicon wafer containing a lens array. 



\subsection{Absorber Design} \label{sec:absorber_design}

\begin{figure}
\centering
\includegraphics[width=\singlecolwidth]{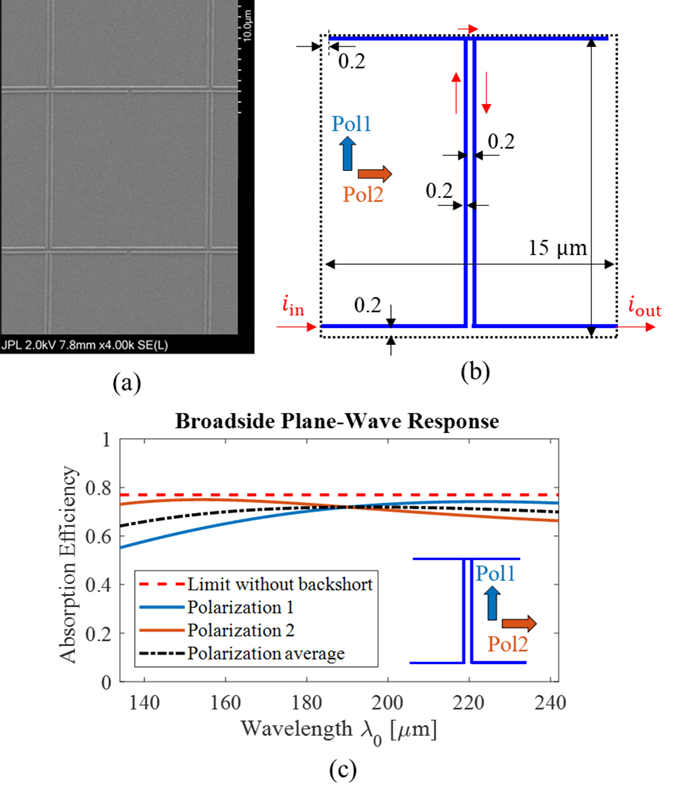}
\caption{\small Optimized $\Pi$-absorber. Orientation is rotated by $90^{\circ}$ with respect to Figure \ref{fig:kid_schematic}. (\textit{a}) SEM picture, (\textit{b}) schematic with dimensions, and (\textit{c}) broadside plane wave response.
\label{fig:absorber_fig}
}
\end{figure}

The aluminum inductor, which is superconducting at microwave frequencies, doubles as an absorber at THz frequencies. A $\Pi$-shaped absorber was optimized for absorbing signals from two orthogonal linear polarizations (Figure \ref{fig:absorber_fig}). When placed in a periodic grid and connected on the sides, this unit cell forms a single meandering inductor. The upper horizontal line of the $\Pi$ does not contribute to the inductance but improves absorption of the vertical polarization via capacitive coupling to the adjacent unit cell. The absorber unit cell is optimized using full-wave simulations considering infinite silicon and air half-spaces above and below the absorber, respectively. To simplify the fabrication and assembly, no quarter-wavelength backshort was used, which limits the potential absorption efficiency to approximately 77\%. The aluminum was modeled with a sheet resistance of $R_s=1.15$ $\Omega/\square$. A minimum line width and line separation of 0.2 {\micron} was enforced, leading to a unit cell periodicity of 15 \micron\ to achieve optimum absorber efficiency. The optimized absorber response for a broadside plane wave coming from the silicon is shown in Figure \ref{fig:absorber_fig}c. A polarization-averaged absorption efficiency of 70\% was achieved over a broad band, close to the theoretical limit of 77\% for an absorber without backshort\footnote{With no backshort the maximum absorption efficiency for normal incidence is $n/(1+n) = 0.77$ for silicon, assuming an index of refraction $n = 3.42$.}. 


\subsection{Lens Design and Optical Efficiency} \label{sec:lens_design}

The lens concentrates incident light from a large area onto a significantly smaller absorber, and greatly impacts the net optical efficiency. Maximizing the optical response of the detector requires minimizing the absorber volume. We have studied square absorbers in a 5x5 or 7x7 unit cell configuration, corresponding to aluminum volumes of 11.3 \micron$^3$ and 20.6 \micron$^3$, respectively. Both configurations use the same unit cell, shown in Figure~\ref{fig:absorber_fig}b. The absorbers are comparable in size to a wavelength in silicon -- 75 {\micron} (=1.3$\lambda$) or 105 {\micron} (=1.8$\lambda$) for $\lambda = 200$ \micron, and therefore low focal ratio ($f/\#$) lenses are needed to minimize the spillover losses \cite{Echternach2022}. The minimum lens $f/\#$ possible was dictated by the minimum wafer thickness we were comfortable handling, which was 300 {\micron} for the detector wafer and 200 {\micron} at the thinnest points of the lens wafer. This leads to an extended hemispherical lens design with $f/\#=0.745$. The simulated spillover efficiency at $\lambda = 200$ {\micron} with respect to the absorbers was 83\% and 90\% for the 5x5 and 7x7 grids, respectively. 

\begin{figure}
\centering
\includegraphics[width=\singlecolwidth]{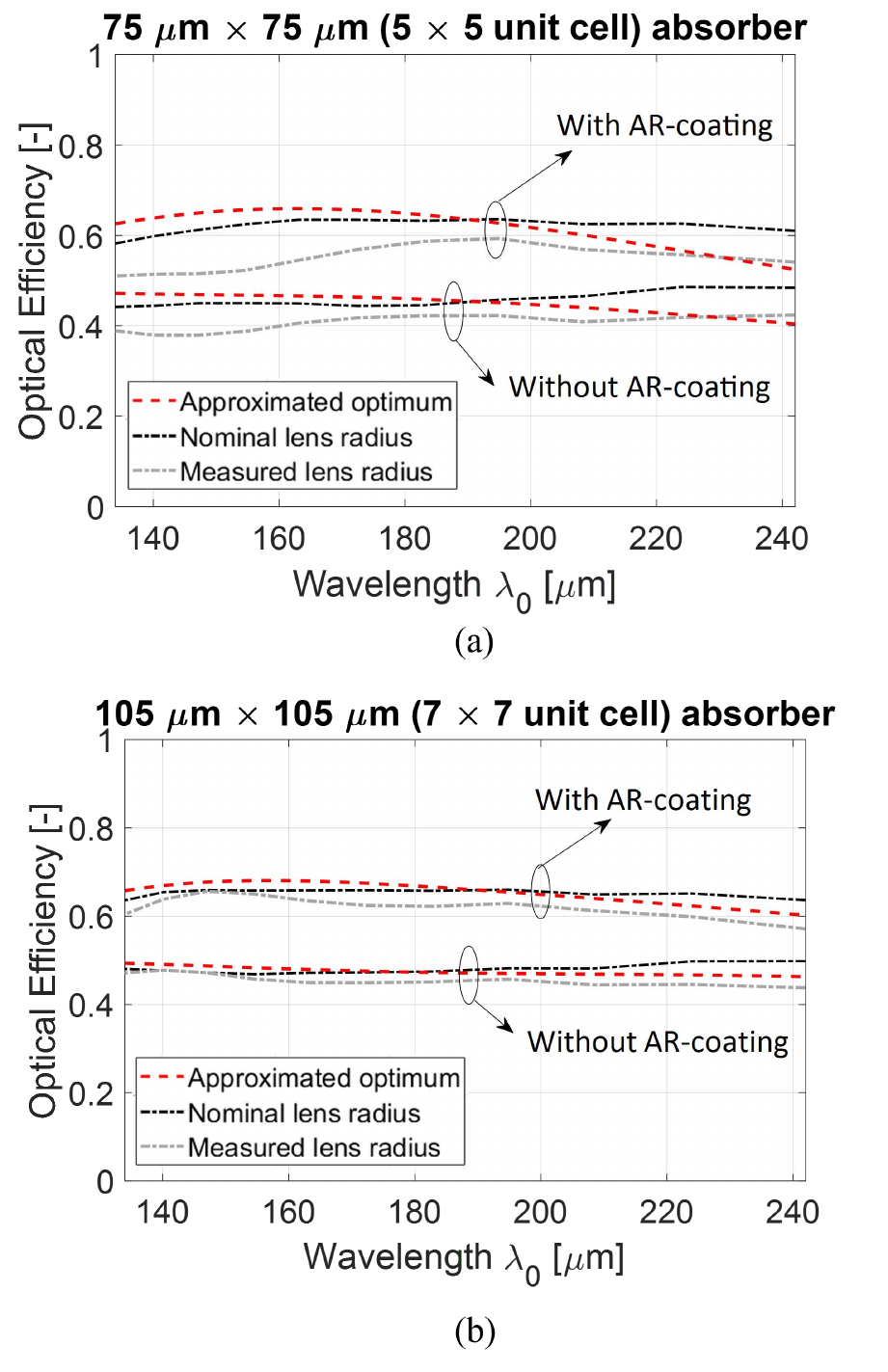}
\caption{\small Simulated optical efficiency of the lens-coupled absorbers, both with and without an AR coating. (\textit{a}) Efficiency with the 5x5 absorber, and (\textit{b}) with the 7x7 absorber. Polarization-averaged efficiencies with the nominal lens radius (\textit{black}) and measured lens radius (\textit{grey}) are compared with the theoretical limits (\textit{red dashed}).
\label{fig:opt_eff_sim}
}
\end{figure}

Figure \ref{fig:opt_eff_sim} shows the simulated optical efficiency of the lens-coupled absorber, assuming a broadside plane wave impinging on the lens surface. Simulations are performed in CST Microwave Studio using the finite-difference time-domain (FDTD) method. The efficiency is analyzed both with and without a Parylene-C ($\epsilon_r=2.62$) quarter-wavelength anti-reflection (AR) coating. The measurements presented in this work are done with lens arrays without an AR-coating. The black lines in Figure \ref{fig:opt_eff_sim} show the polarization-averaged efficiency, which is between $44\%-48\%$ for the 5x5 absorber, and between $47\%-50\%$ for the 7x7 absorber. 

The red dashed lines in Figure \ref{fig:opt_eff_sim} indicate the approximate theoretical limits to the optical efficiency, including the absence of a backshort ($\sim$$77$\% for a broadside plane wave), absence of an AR-coating ($\sim$$70$\% limit without coating\footnote{With no AR-coating the transmission efficiency from free space into the silicon lens for a normally-incident plane wave is $1 - (1-n)^2/(1+n)^2 = 0.70$, assuming an index of refraction $n = 3.42$.}), and the simulated spillover of the focal plane field. This spillover efficiency is a function of wavelength and is calculated as the ratio of the power that is geometrically available to the finite sized absorber to the total power incident in the lens focal plane. These calculations are performed using an efficient quasi-analytical method \cite{LLombart2015}. At long wavelengths the absorbers become electrically small and the effective area can be larger than the geometrical area \cite{LLombart2018}, such that the spillover efficiency as calculated here underestimates the true absorption efficiency. This effect is responsible for the drop in the theoretical absorption limits shown in Figure \ref{fig:opt_eff_sim}, which fall below the simulated efficiencies at long wavelengths, particularly for the smaller 5x5 unit cell absorber. Additionally, the approximated optimum is pessimistic, as an absorber without backshort under a low $f/\#$ lens becomes more sensitive to plane waves coming from angles beyond the silicon-air critical angle.

\begin{figure}
\centering
\includegraphics[width=\singlecolwidth]{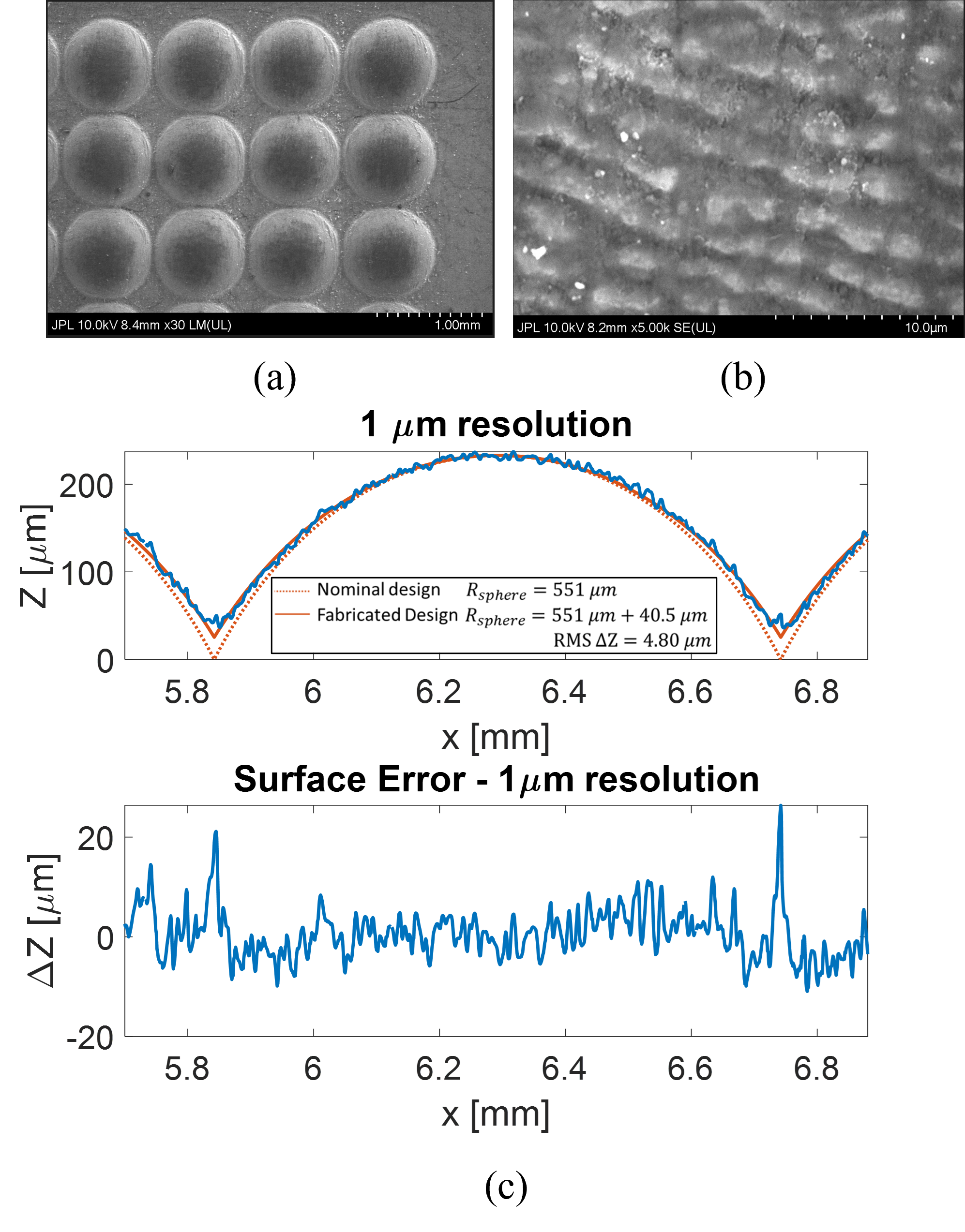}
\caption{\small  Machined 21x21 lens array with low $f/\#$ lenses. (\textit{a}) SEM of some lenses, (\textit{b}) close-up SEM showing surface roughness, and (\textit{c}) zoom-in of designed, measured, and simulated lens profile.
\label{fig:lens_profile}
}
\end{figure}

The 21x21 lens array used in this measurement was machined with laser ablation technology at Veldlaser\footnote{\url{https://www.veldlaser.nl/en/}}. SEM pictures of a portion of the array and zoom-in of the lens surface are shown in Figure \ref{fig:lens_profile}. The surface profile of a full row of lenses (20 mm line scan with a 1 {\micron} resolution) was measured using a Mitaka NH-5NS surface mapping tool\footnote{\url{https://www.mitakakohki.co.jp/english/industry/nh_series/lineup/nh-5ns-.html}}, indicating an increased (by 40.5 {\micron}) lens radius, and a surface roughness rms of 4.8 {\micron}. We repeated the full-wave simulations with the larger lens radius and found a reduction in the optical efficiency of up to 10\% (Figure~\ref{fig:opt_eff_sim}). The impact of the 4.8 {\micron} rms surface roughness was estimated using the standard Ruze efficiency, adapted for refractive lenses\cite{Ruze1966,Buttgenbach1993}, which suggests an additional penalty of 12\%. As a result of the larger radius and the finite surface roughness, the wavelength-averaged optical efficiency of the 7x7 unit cell absorber is simulated to be $40\%$.



In parallel with these initial measurements, NASA's Goddard Space Flight Center (GSFC) has developed 1008-pixel microlens arrays using etching via greyscale lithography. These arrays show excellent surface quality even for PRIMA's shortest wavelengths, as well as a short manufacture time \cite{cothard_2024ApOpt}. They are now being manufactured in the PRIMA flight format, and are the baseline plan for PRIMA's KIDs.

\section{Experimental Setup} \label{sec:experimental_setup}

\begin{figure}
\centering
\includegraphics[width=\singlecolwidth]{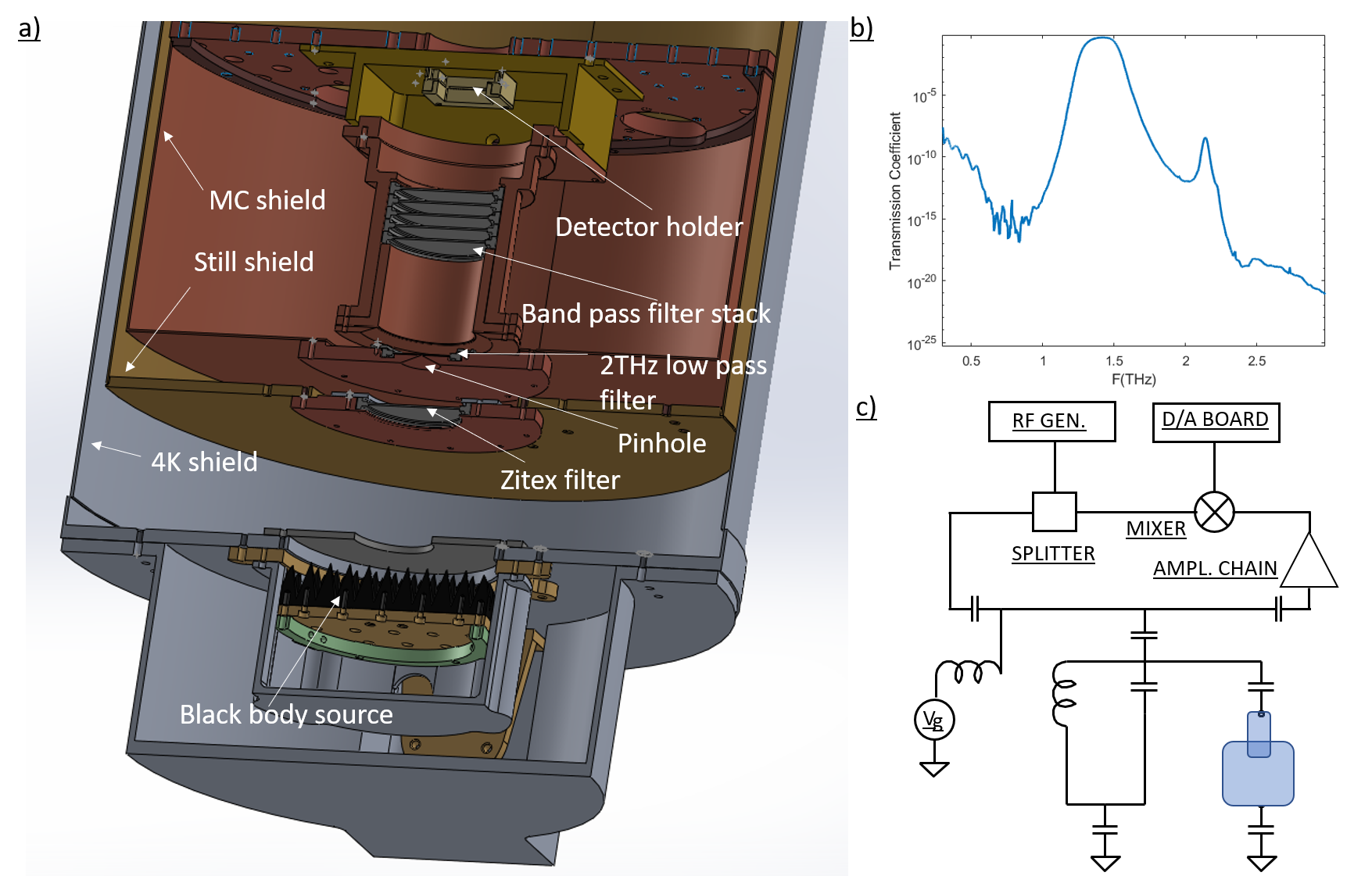}
\caption{\small (\textit{a}) Optical setup. (\textit{b}) Filter stack transmission as a function of frequency. Peak transmission is 44\% at 1.42 THz, and FWHM is 0.17 THz. (\textit{c}) Readout schematic.
\label{fig:cryo_bb_fig}
}
\end{figure}

Figure \ref{fig:cryo_bb_fig} shows the optical setup. Radiation is generated by a variable temperature blackbody source that is mounted to the exterior of a cryogen-free dilution refrigerator’s 4 K shield with low-thermal-conductivity flexures. The blackbody is thermally connected to the 4 K stage through copper braids chosen to provide a heat conductance that balances blackbody speed against thermal loading of the shield. A flange with a 3.8~cm opening limits the amount of blackbody radiation to avoid overheating the still shield. Radiation enters the still shield through a 4.6~cm diameter opening covered with a Zitex filter, and the mixing chamber shield through a 500 {\micron} diameter aperture. This small aperture reduces the optical throughput, which allows the desired optical power range ($3\times10^{-21} - 1\times 10^{-15}$ W) to be reached with the available blackbody temperatures ($4-33$ K). Additionally, the diffraction introduced by the $500$ \micron\ aperture ($=2.4\lambda$ at 210 $\mu$m) causes a reasonably constant illumination level over the detector array. The aperture flange holds behind it a 2 THz low-pass filter. The detector array is located close to the mixing chamber plate inside a copper housing, which also secures in place a stack of five bandpass filters centered at 1.41 THz (213 \micron). The peak transmission of the full filter stack is 44\%, and the FWHM is 0.17 THz (Figure \ref{fig:cryo_bb_fig}b). Both the still and mixing chamber shields are coated with a mixture of black epoxy and carbon black\cite{Persky1999}. The 500 $\mu$m aperture and the bandpass filter stack are sunk at the 125mK temperature of the mixing chamber, so thermal emission from these sources is expected to be negligible.

The blackbody radiation incident on the front of the lens is calculated as\cite{Echternach2022}:
\begin{equation} \label{eq:bb_integration}
P_\mathrm{inc} = \Omega_\mathrm{lens} A_\mathrm{ap} \eta_\mathrm{ap} \int B_\nu(\nu,T_\mathrm{BB}) \, T(\nu) \, d\nu,
\end{equation}

\noindent where $B_\nu(\nu,T_\mathrm{BB})$ is the Planck function at the blackbody temperature $T_\mathrm{BB}$, $T(\nu)$ is the transmission through the filter stack shown in Figure \ref{fig:cryo_bb_fig}b, $\Omega_\mathrm{lens}$ is the solid angle subtended by the lens as seen from the 500 $\mu$m pinhole aperture, and $A_\mathrm{ap}$ is the aperture area. The $\eta_\mathrm{ap}$ parameter in Equation \ref{eq:bb_integration} accounts for the loss in incident power due to the small size of the pinhole, combined with the finite angular size of the blackbody as viewed from the pinhole aperture. We calculate $\eta_\mathrm{ap}$ by considering a plane wave normally incident on the pinhole in the time-reverse direction, and computing the fraction of the far-field radiation that couples to the blackbody\cite{Blevin1970}. Modeling the pinhole as a circular hole with radius $a$ in a thin, perfectly conducting sheet, a vector diffraction calculation gives the transmitted power per unit solid angle as\cite{jackson_classical_1999}:
\begin{equation} \label{eq:diff_pattern}
\frac{dP}{d\Omega} = P_i \frac{(ka)^2}{4\pi}\big(\mathrm{cos}^2(\theta) + \mathrm{cos}^2(\phi) \,\mathrm{sin}^2(\theta)\big) \bigg| \frac{2J_1\big(ka\,\mathrm{sin}({\theta})\big)}{ka\,\mathrm{sin}({\theta})}\bigg|^2,
\end{equation}

\noindent where $P_i$ is the power normally incident on the aperture. The 3.8 cm opening in front of the blackbody subtends a $21.3^{\circ}$ angle as viewed from the pinhole. Integrating the diffraction pattern over this solid angle yields $\eta_\mathrm{ap} = 0.38$.


As shown in Figure \ref{fig:cryo_bb_fig}c, measurements are performed by sending a microwave signal at the resonant frequency of the KID through a feedline that is capacitively coupled to the resonator. The transmitted signal is amplified by a cryogenic amplifier and a chain of warm amplifiers before being down-converted by an in-phase quadrature (IQ) mixer. The down-converted in-phase and quadrature signals are digitized with a sample rate of 100kHz and saved separately.

\section{Measurements} \label{sec:measurements}

We cryogenically tested a wafer patterned with 14 large volume KIDs (with 7x7 unit cell inductors) and 14 small volume KIDs (with 5x5 unit cell inductors). The IDCs and coupling capacitors of these KIDs were designed to produce resonance frequencies in the $2-3$ GHz range, and coupling quality factors of $Q_c = 10^4$. A total of 7 large volume KIDs were identified, and all of these showed a response to the cryogenic blackbody. We chose the resonator at $f_r = 2.1$ GHz with the strongest optical response for detailed study. 

While only 7 of the 14 large volume KIDs on this wafer were identified, subsequent improvements to the fabrication procedure have increased this yield. The fabrication process employed on this wafer was designed for liftoff of small Al/AlO$_\mathrm{x}$/Al junctions, and not for a grid of closely spaced lines. We have improved the yield by 1) individually characterizing the bilayers in the liftoff process to input into 3D proximity correction software, along with the nominal grid design, for e-beam lithography exposure, and by 2) improving control of the undercut of the bottom layer. The fabrication yield on more recent large format arrays is $>90$\%\cite{foote_ltd2023}.

\subsection{Dark Measurements} \label{sec:dark_measurements}

We begin by characterizing the detector with the cryogenic blackbody set to 5.4K, which corresponds to a negligible incident power of 0.027 aW at the front of the lens, effectively placing the detector in the dark. We perform a frequency sweep of the microwave source over the resonance with a drive power of $P_g = -98$ dBm at the detector, which is approximately 6~dB below the bifurcation power\cite{Swenson2013}. We fit the system-corrected microwave transmission using:
\begin{equation} \label{eq:s21_basic}
S_{21} = 1 - \frac{Q_r/Q_c \, (1 + j\,\mathrm{tan}(\phi))}{1 + 2 j Q_r x},
\end{equation}

\noindent where $x = (f - f_r)/f_r$ is the fractional detuning away from the resonator frequency $f_r$, $Q_r$ is the resonator quality factor, $Q_c$ is the coupling quality factor, and the $\mathrm{tan}(\phi)$ term accounts for impedance mismatches along the transmission line \cite{zmuidzinas2012,Khalil2012}. At $T=125$ mK this fit yields $Q_r = 10^4$ and $Q_c = 1.1 \times 10^4$, close to the target value.

\begin{figure}
\centering
\includegraphics[trim = 60 280 60 50, width=\singlecolwidth]{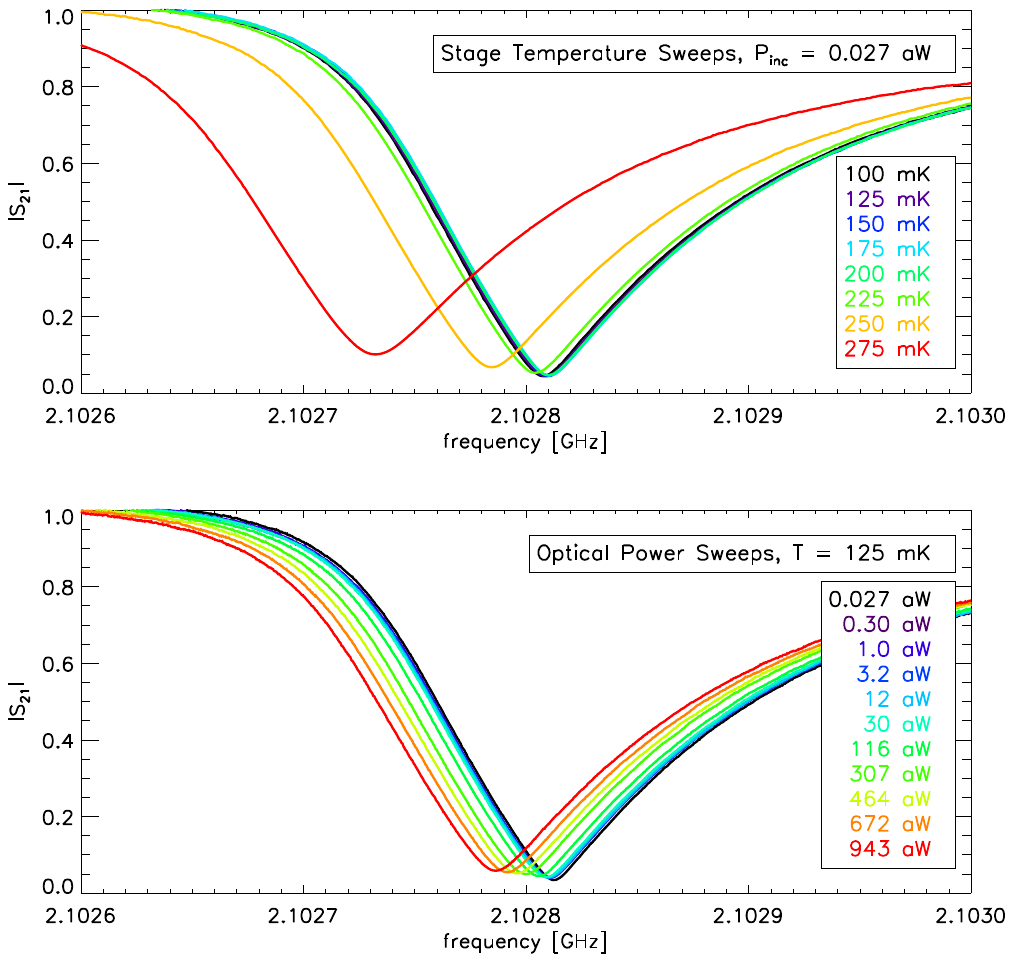} 
\caption{\small System-corrected microwave transmission profiles as a function of stage temperature for fixed optical loading (\textit{top}), and as a function of optical loading at fixed stage temperature (\textit{bottom}).
\label{fig:reso_shift_comparison}
}
\end{figure}
\begin{figure}
\centering
\includegraphics[trim = 60 520 60 50, width=\singlecolwidth]{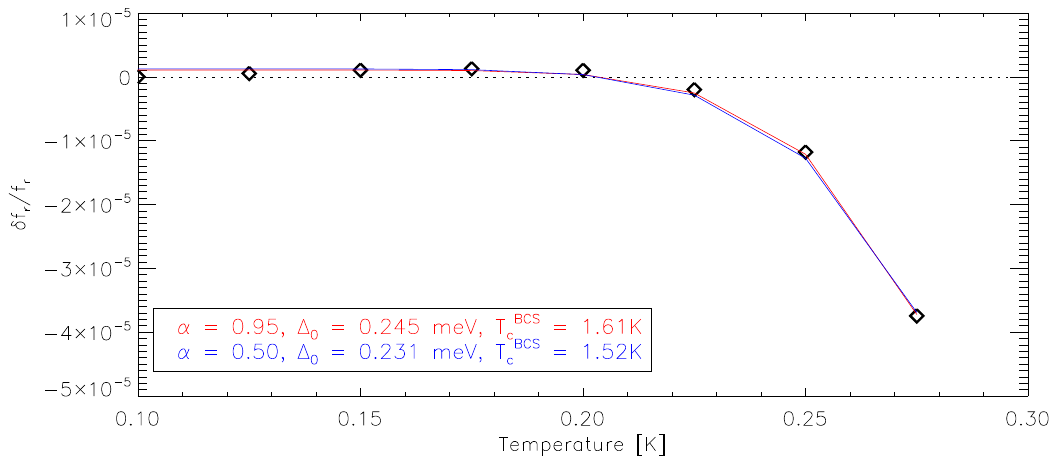} 
\caption{\small Fractional frequency shift vs. stage temperature for a dark measurement, along with fits to $x = \delta f_r/f_r$ for a range of fixed $\alpha$ and variable $\Delta_0$. Models are $x = x_\mathrm{MB} + \delta x$ for a zero temperature offset $\delta x$. The $T_c$ is computed following the BCS relation $\Delta_0 = 1.76 k_B T_c$. 
\label{fig:reso_shift_dark}
}
\end{figure}

Transmission profiles as a function of the stage temperature for $T = 100 - 275$ mK are shown in Figure \ref{fig:reso_shift_comparison}, and the estimated fractional frequency shift ($x = \delta f_r/f_r$) is shown in Figure \ref{fig:reso_shift_dark}. We fit this data with the standard Mattis-Bardeen expression\cite{zmuidzinas2012}:
\begin{equation}
x_\mathrm{MB} = -\frac{\alpha \gamma S_2(\omega)}{4 N_0 \Delta_0} n_\mathrm{qp},
\label{eq:x_mb}
\end{equation}


\noindent where $\alpha$ is the kinetic inductance fraction, $\Delta_0$ is the gap energy, and $\gamma = 1$ is appropriate for the thin films used here. We adopt a density of states of $N_0 = 1.72 \times 10^{10}$ $\mu$m$^{-3}$\,eV$^{-1}$ \cite{Gao2008JLTP}. The quasiparticle number density $n_\mathrm{qp}$ is given by $n_\mathrm{th} = 2 N_0 \sqrt{2 \pi k_B T \Delta_0} \, \mathrm{exp}(-\Delta_0/k_B T)$, and we also use the standard expression for $S_2(\omega)$ appropriate for a thermal quasiparticle distribution\cite{zmuidzinas2012}:
\begin{equation}
S_2(\omega) = 1 +  \sqrt{\frac{2\Delta_0}{\pi k_B T}} \mathrm{exp}\bigg(-\frac{\hbar\omega}{2k_BT}\bigg)I_0\bigg(\frac{\hbar\omega}{2k_BT}\bigg),
\end{equation}

\noindent where $\omega = 2\pi f$. When using Equation \ref{eq:x_mb} to model the data shown in Figure \ref{fig:reso_shift_dark} we allow $\alpha$ and $\Delta_0$ to vary, as discussed below. 

The two models shown in Figure \ref{fig:reso_shift_dark} are fits to $x(T)$ with $\alpha$ fixed at 0.95 and 0.50, respectively, and $\Delta_0$ treated as a free parameter. For this temperature range $\alpha$ and $\Delta_0$ are almost entirely degenerate, and additional information must be used to constrain these two parameters. Given the narrow 0.2 {\micron} line width the kinetic inductance in the absorber is expected to be significantly larger than the geometric inductance, and electromagnetic simulations of the absorber with Sonnet\footnote{https://www.sonnetsoftware.com/} indicate $\alpha \approx 0.95$. We thus adopt $\alpha = 0.95$ and $\Delta_0 = 0.245$ meV as our preferred model. Assuming the standard BCS relation between the gap energy and transition temperature ($\Delta_0 = 1.76 k_B T_c$) this corresponds to $T_c = 1.61$ K.

With $\alpha = 0.50$ the best fit value of $\Delta_0$ is $0.231$ meV, only $\approx$$6\%$ lower than in our preferred model. This value of $\alpha$ would be significantly lower than expected from the simulations, and also lower than the values of $\alpha$ obtained in other narrow line aluminum inductors \cite{McCarrick2014, haileydunsheath2018, Janssen_2023}. We consider this a conservative lower limit to the kinetic inductance fraction, and $\Delta_0 = 0.231$ meV a corresponding lower limit to the gap energy.

We note that while these estimates of $\alpha$ and $\Delta_0$ contain some uncertainty, precise values of these parameters are not required to measure the detector sensitivity. The calculation of the detector NEP and optical efficiency presented in Section \ref{sec:nep} does not require any knowledge of $\alpha$, and is only weakly dependent on the adopted value of $\Delta_0$ through Equation \ref{eq:nep_gamma}. We discuss this dependence more fully in Section \ref{sec:nep}.

\subsection{Optical Measurements} \label{sec:optical_measurements}

\begin{figure}
\centering
\includegraphics[trim = 40 400 80 50, width=\singlecolwidth]{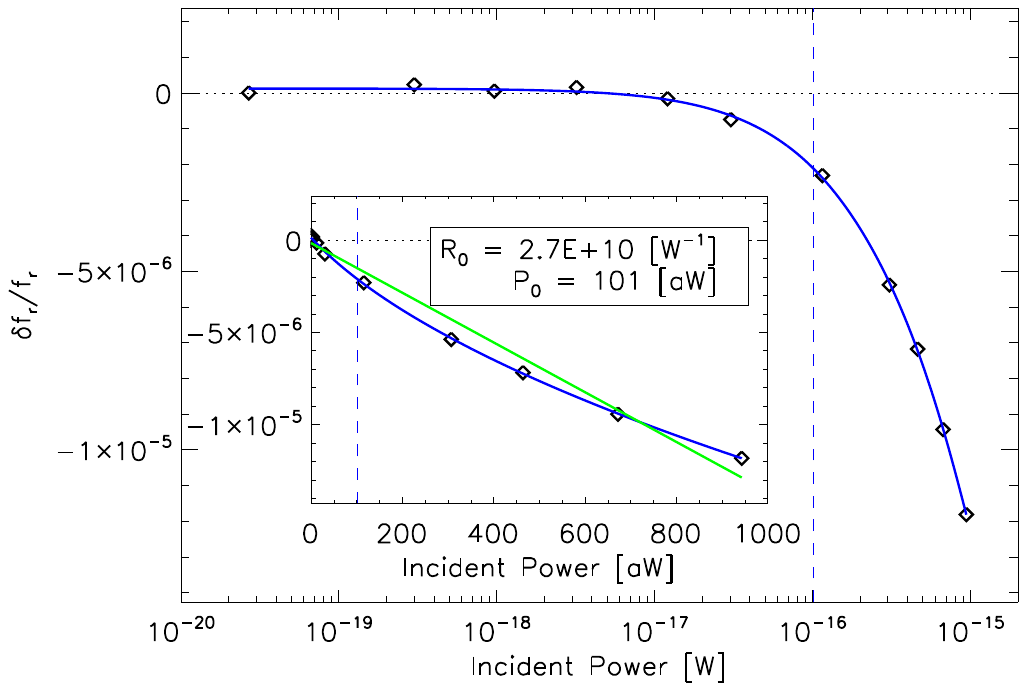} 
\caption{\small Fractional frequency shift vs. incident optical power, along with a fit (\textit{blue}) for Equation \ref{eq:resp_integrated}. Vertical dashed line shows the fitted value of the parameter $P_0$. Inset shows the same on a linear scale, along with the result of a fit for a constant responsivity (\textit{green}) for comparison.
\label{fig:reso_shift_optical}
}
\end{figure}

We measure the optical response of this KID by fixing the stage temperature at 125 mK and running the cryogenic blackbody between $5.4-33.4$ K, corresponding to an incident power range of $P_\mathrm{inc} = 0.027 - 943$ aW. The resultant resonator profiles are shown in the bottom panel of Figure \ref{fig:reso_shift_comparison}, and the fractional frequency shift as a function of optical loading is shown in Figure \ref{fig:reso_shift_optical}.



If the quasiparticle generation rate due to the absorption of microwave readout power may be neglected, the change in the fractional frequency shift with respect to the change in incident power ($P_\mathrm{inc}$) may be written as\cite{zmuidzinas2012}:
\begin{subequations} \label{eq:resp_exp}
\begin{align} 
\frac{dx}{dP_\mathrm{inc}}
&= \eta_\mathrm{opt} \frac{\alpha \gamma S_2(\omega)}{4 N_0 \Delta_0} \frac{\eta_\mathrm{pb} \tau_\mathrm{max}}{\Delta_0 V} \\
&~~~~~~~\bigg[\bigg(1 + \frac{n_\mathrm{th}}{n^*}\bigg)^2 + \frac{2 \eta_\mathrm{pb} \eta_\mathrm{opt} P_\mathrm{inc} \tau_\mathrm{max}}{n^* \Delta_0 V} \bigg]^{-1/2} \nonumber \\
&= R_0 \bigg[1 + \frac{P_\mathrm{inc}}{P_0} \bigg]^{-1/2},
\end{align}
\end{subequations}

\noindent where we have introduced:
\begin{subequations} \label{eq:AB_def}
\begin{align} 
R_0 &= \eta_\mathrm{opt} \frac{\alpha \gamma S_2(\omega)}{4 N_0 \Delta_0} \frac{\eta_\mathrm{pb} \tau_\mathrm{max}}{\Delta_0 V} \bigg(1 + \frac{n_\mathrm{th}}{n^*}\bigg)^{-1} \\
P_0 &= \frac{n^* \Delta_0 V}{2 \eta_\mathrm{pb} \eta_\mathrm{opt} \tau_\mathrm{max}} \bigg(1 + \frac{n_\mathrm{th}}{n^*}\bigg)^2.
\end{align}
\end{subequations}

\noindent In these equations $\eta_\mathrm{opt}$ is the optical efficiency of the lens and absorber, such that the power absorbed in the detector is $P_\mathrm{abs} = \eta_\mathrm{opt} P_\mathrm{inc}$. The quantity $\eta_\mathrm{pb}$ is the pair-breaking efficiency, defined as the fraction of the photon energy that breaks Cooper pairs, and $V$ is the inductor volume. We adopt the common empirical description of the quasiparticle lifetime: $\tau_\mathrm{qp} = \tau_\mathrm{max} (1 + n_\mathrm{qp}/n^*)^{-1}$, where $n_\mathrm{qp}$ is the quasiparticle number density at the given optical load, and $\tau_\mathrm{max}$ and $n^*$ are constants determined by the film properties.

In Equation \ref{eq:AB_def}, $P_0$ is constant with incident power and $R_0$ depends on the optical loading only through the parameter $S_2(\omega)$, which depends on the quasiparticle energy distribution. We estimate that $S_2(\omega)$ varies by only $\approx$$10\%$ for the range of optical loading considered here. Neglecting this variation and treating $R_0$ as a constant, we may integrate Equation \ref{eq:resp_exp} to find:
\begin{equation}
x(P_\mathrm{inc}) - x(P_\mathrm{inc}=0) = 2 R_0 P_0 \bigg[ \bigg(1 + \frac{P_\mathrm{inc}}{P_0} \bigg)^{1/2} - 1\bigg].
\label{eq:resp_integrated}
\end{equation}

\noindent A fit to the data in Figure \ref{fig:reso_shift_optical} with this approach yields $R_0 = 2.7 \times 10^{10}$ W$^{-1}$ and $P_0 = 101$ aW (see Figure \ref{fig:reso_shift_optical} inset).





\begin{figure}
\centering
\includegraphics[trim = 50 60 60 60, width=\singlecolwidth]{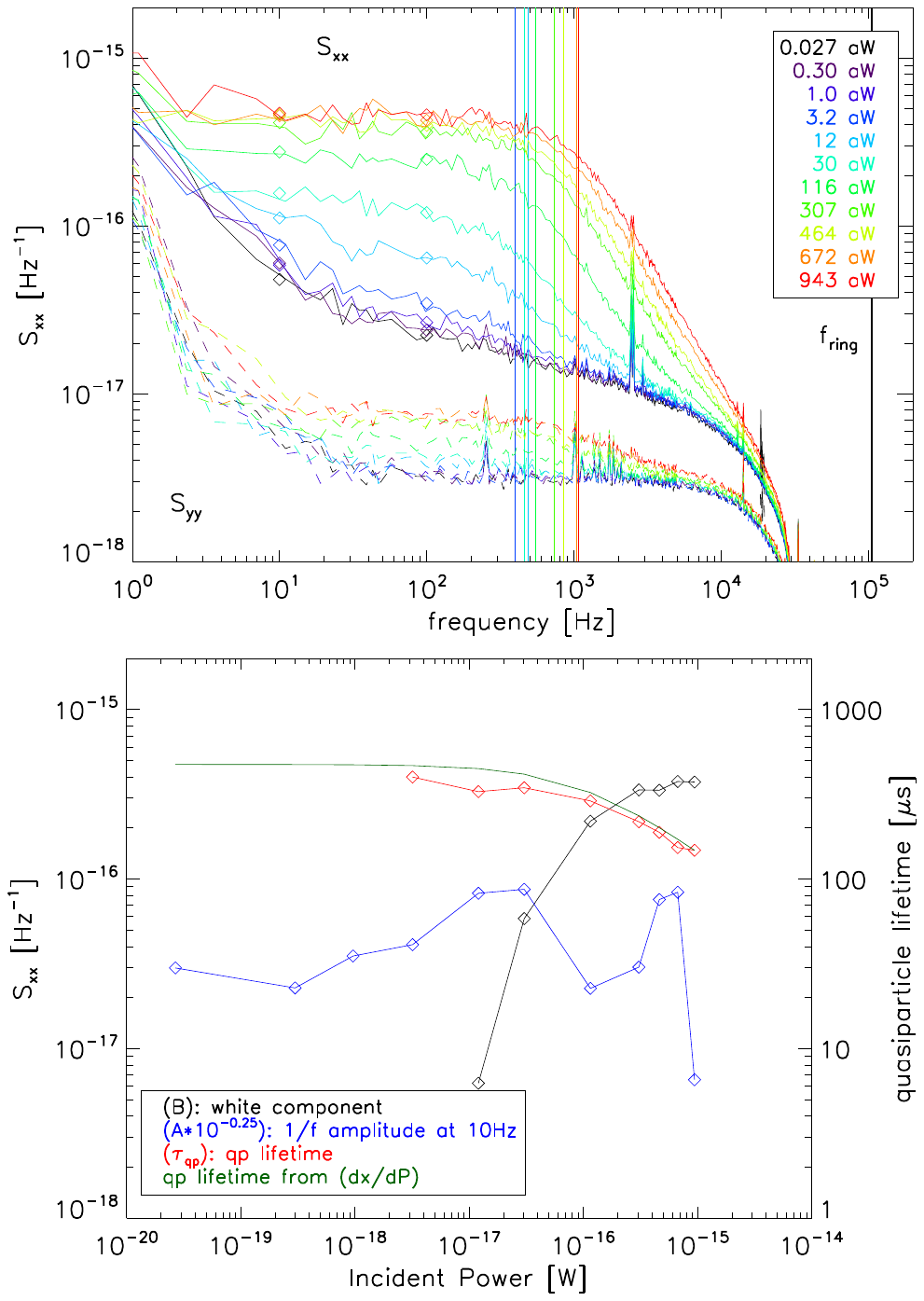} 
\caption{\small (\textit{top}) Fractional frequency noise over a range of optical loads, with electrical noise spikes at 60 Hz and harmonics removed. Noise measured orthogonal to the phase direction shown as dashed lines ($S_\mathrm{yy}$). Vertical lines indicate the fitted roll-off of the resonator component, and thick line at 100 kHz indicates the resonator ring-down frequency. (\textit{bottom}) Fitting results of the PSDs using Equation \ref{eq:sxx_form}. The quasiparticle lifetime estimated from the shape of $dx / dP_\mathrm{inc}$ is overplotted (\textit{green}).
\label{fig:optical_noise}
}
\end{figure}

At each optical load we obtain a 20 second time stream of IQ data at fixed tone frequency, and calculate the fractional frequency noise power spectral density (PSD), $S_\mathrm{xx}$, shown in Figure \ref{fig:optical_noise}. Focusing on sampling frequencies $\gt$$10$ Hz, we see at low optical loading the PSD is dominated by a power law component consistent with two-level system (TLS) noise\cite{Gao2008TLSmodel}, while at high optical loading the white photon noise dominates. The white noise component rolls off with a knee frequency of $\approx$$0.4-1$ kHz, which is well below the $\approx$$100$ kHz resonator ring-down frequency ($f_\mathrm{ring} = f_r/2Q_r$), or the $\approx$$50$ kHz roll-off due to the warm electronics filter, and we attribute this knee to the quasiparticle lifetime. We fit the resonator contribution to the measured PSD at $f > 10$ Hz using:
\begin{equation} \label{eq:sxx_form}
S_\mathrm{xx} = \frac{A f^{-0.25} + B}{1 + (2 \pi f \tau_\mathrm{qp})^2} + \frac{C}{1 + (2 \pi f \tau_\mathrm{c})^2},
\end{equation}

\noindent where $A$ represents the amplitude of the $1/f$ term, which we find is well described by an $f^{-0.25}$ power law, and $B$ is the white noise term. We find evidence for a second white noise component in excess of the electronics noise, and account for this with the second term in Equation \ref{eq:sxx_form}. Typical values here are $C = 1 \times 10^{-17}$ Hz$^{-1}$, and $\tau_\mathrm{c} = 40$ $\mu$s (corresponding to a knee frequency $f_c = 4$ kHz). The bottom panel of Figure \ref{fig:optical_noise} shows the fitted white noise amplitude ($B$), the $1/f$ noise amplitude at 10 Hz ($A \times 10^{-0.25}$), and $\tau_\mathrm{qp}$. At incident power levels above 3 aW there is sufficient photon noise to obtain a fit to $\tau_\mathrm{qp}$, and we estimate $\tau_\mathrm{qp}$ decreases from 400 $\mu$s at 3.1 aW to 148 $\mu$s at 943 aW. 

The frequency response to optical power may be used to provide an additional constraint on the quasiparticle lifetime. With the assumption that the changing slope $dx / dP_\mathrm{inc}$ seen in Figure \ref{fig:reso_shift_optical} is due primarily to the change in $\tau_\mathrm{qp}$, we expect that the two quantities are proportional. Making use of Equation \ref{eq:resp_exp}, we may express this as $\tau_\mathrm{qp} \propto [1 + P_\mathrm{inc}/P_0]^{-1/2}$. Normalizing this equation to the absolute value of $\tau_\mathrm{qp} = 148$ $\mu$s at $P_\mathrm{inc} = 943$ aW measured from fitting the PSD roll-off, and adopting $P_0 = 101$ aW from the analysis of the data in Figure \ref{fig:reso_shift_optical}, we construct a model for $\tau_\mathrm{qp}$ as a function of $P_\mathrm{inc}$. This is shown as the green curve in Figure \ref{fig:optical_noise} (bottom). We find that this model is consistent with the values of $\tau_\mathrm{qp}$ estimated by fitting the PSD roll-off, and indicates that $\tau_\mathrm{qp}$ saturates at $400-500$ $\mu$s as the optical load drops below 1 aW.


\subsection{Noise Equivalent Power} \label{sec:nep}

With the electrical noise and response to incident power measured, we may compute the NEP referenced to power incident on the front of the lens as:
\begin{equation} \label{eq:nep_def}
\mathrm{NEP_\mathrm{inc}} = \sqrt{S_\mathrm{xx}} \bigg(\frac{dx}{dP_\mathrm{inc}}\bigg)^{-1} \sqrt{1 + (2 \pi f \tau_\mathrm{qp})^2}.
\end{equation}

\noindent In Section \ref{sec:optical_measurements} we used a detailed detector model to develop an expression for $dx/dP_\mathrm{inc}$, but then grouped terms to arrive at a simple 2-parameter model of $x(P_\mathrm{inc})$ (Equations \ref{eq:resp_exp}, \ref{eq:AB_def}, and \ref{eq:resp_integrated}). This model provides an excellent match to the measurements (Figure \ref{fig:reso_shift_optical}), and we consider the resulting description of $dx/dP_\mathrm{inc}$ to be comparable to what would be obtained from an interpolation or other purely phenomenological description of this data. We use the measured $S_\mathrm{xx}$ in Equation \ref{eq:nep_def}, and consider the first two terms of this equation to be directly obtained from the data, with minimal influence of any model assumptions.

\begin{figure}
\centering
\includegraphics[trim = 60 400 70 60, width=\singlecolwidth]{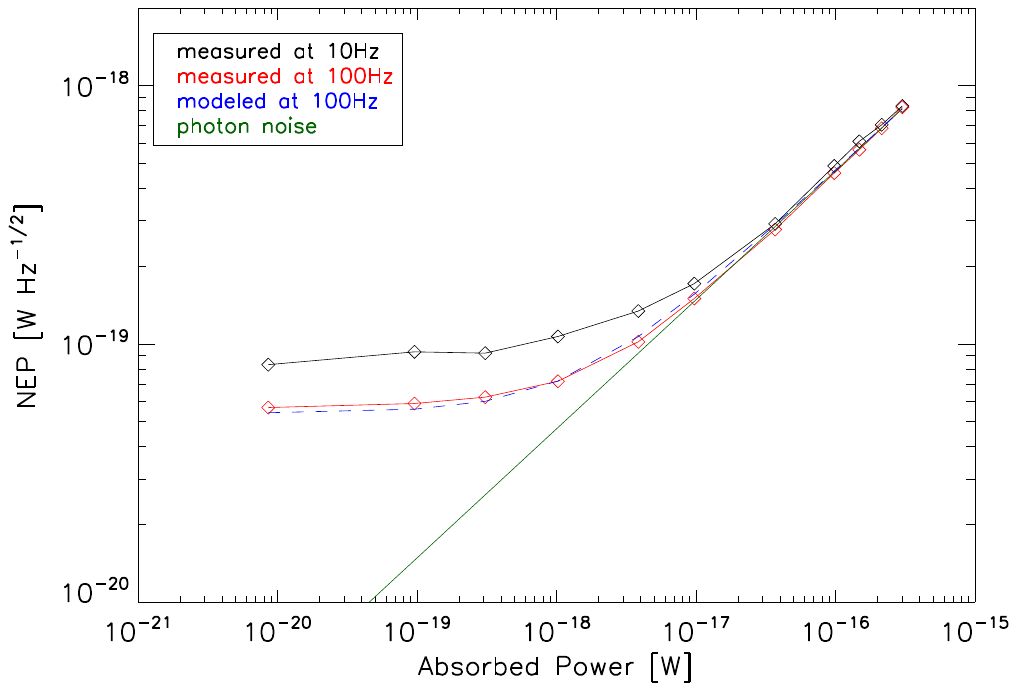} 
\caption{\small NEP referenced to absorbed power at sampling frequencies of 10 Hz (\textit{black}) and 100 Hz (\textit{red}), for an estimated optical efficiency of $\eta_\mathrm{opt} = 0.32$. Overplotted is the photon NEP (\textit{green}) and a fit to the measured 100 Hz noise combining the photon noise and a fixed contribution added in quadrature (\textit{blue dashed}).
\label{fig:NEPA}
}
\end{figure}

\begin{figure}
\centering
\includegraphics[trim = 60 400 70 60, width=\singlecolwidth]{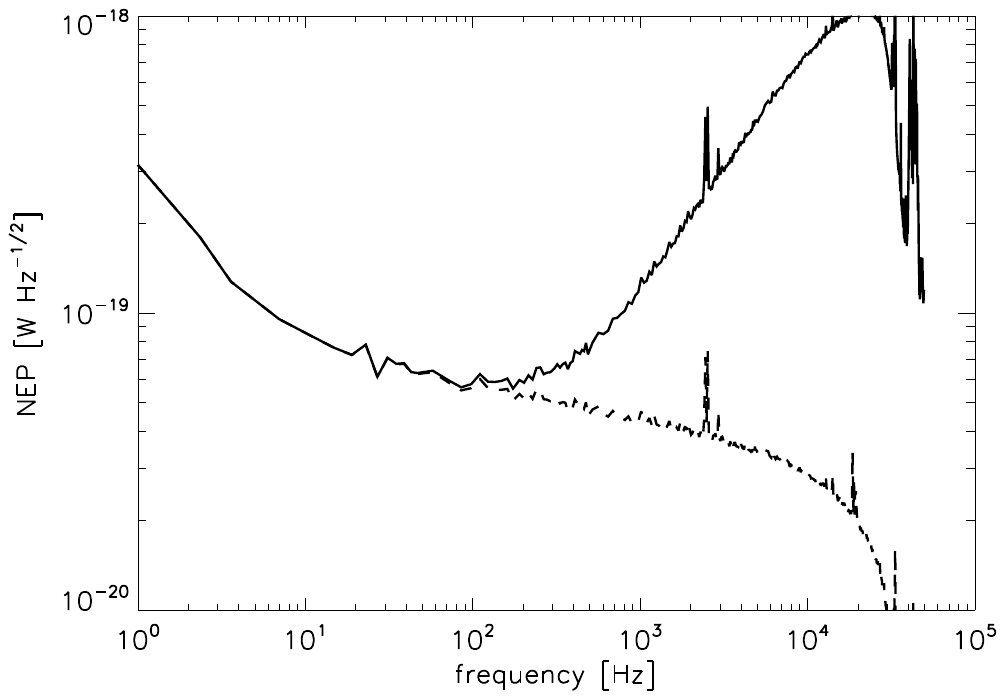} 
\caption{\small NEP referenced to absorbed power as a function of sampling frequency measured for an absorbed power $P_\mathrm{abs} = 0.0086$ aW, both before correcting for the reduced response at high frequency (\textit{dashed}), and after the correction, assuming $\tau_\mathrm{qp} = 415$ $\mu$s (\textit{solid}).
\label{fig:nep_spectrum}
}
\end{figure}

The third term in Equation \ref{eq:nep_def} describes the reduced responsivity at high sampling frequencies due to the finite $\tau_\mathrm{qp}$. As described in Section \ref{sec:optical_measurements} this parameter is obtained through fits to the noise PSDs, supplemented by an extrapolation of the shape of $dx/dP_\mathrm{inc}$ to low optical powers. We adopt a value of $\tau_\mathrm{qp} = 415$~$\mu$s in the low loading limit, but acknowledge some uncertainty in this number. Changes to this $\tau_\mathrm{qp}$ will have little impact on the NEP at sampling frequencies of $10-100$ Hz, and hence on the inferred optical efficiency (Figure \ref{fig:NEPA}), but will determine how quickly the NEP degrades at higher frequencies (Figure \ref{fig:nep_spectrum}).

At high optical loading our noise is dominated by photon generation-recombination (GR) noise, and we may use the method developed by Ref. \citenum{janssen2013} to compute the optical efficiency $\eta_\mathrm{opt}$, and with it the NEP referenced to absorbed power. For background limited operation the photon-limited NEP is:
\begin{equation} \label{eq:nep_gamma}
\mathrm{NEP^2_\gamma} = 2 h \nu P_\mathrm{abs} \bigg(1 + n_0 + \frac{2\Delta_0}{h \nu \eta_\mathrm{pb}} \bigg),
\end{equation}

\noindent where the first term is the photon (generation) shot noise, the second term represents wave noise (for a photon occupation number $n_0$ in the detector), and the last term represents recombination noise. The wave noise term is negligible at these frequencies. For $\Delta_0 = 0.245$ meV and $\eta_\mathrm{pb} = 0.5$ (Section \ref{sec:pairbreak}), this last term is $2\Delta_0 / h \nu \eta_\mathrm{pb} = 0.17$, indicating that recombination noise is a small contribution. The ratio of this expression for $\mathrm{NEP}_\gamma$ to the measured $\mathrm{NEP}_\mathrm{inc}$ in the photon noise limited regime is equal to $\sqrt{\eta_\mathrm{opt}}$. 

In Figure \ref{fig:NEPA} we show the result of this analysis. Based on the photon noise comparison we estimate an optical efficiency of $\eta_\mathrm{opt} = 32\%$. When referenced to absorbed power, the NEP at 10 Hz and 100 Hz sample frequencies approach $9 \times 10^{-20}$ \nepunits\ and $6 \times 10^{-20}$ \nepunits, respectively, in the low loading limit. Using a smaller value of $\Delta_0 = 0.231$ meV (Section \ref{sec:dark_measurements}) will slightly reduce the recombination noise term in Equation \ref{eq:nep_gamma}, but will only reduce the inferred optical efficiency by $\sim$$2\%$. In Figure \ref{fig:nep_spectrum} we show the NEP spectrum measured under the lowest optical loading ($P_\mathrm{abs} = 0.0086$ aW) as a function of sample frequency, calculated from Equation \ref{eq:nep_def} with $\tau_\mathrm{qp} = 415$~$\mu$s. 

The 32\% measured optical efficiency is slightly lower than the 40\% absorption efficiency simulated for the 7x7 unit cell absorber and the as-built lens (Figure \ref{fig:opt_eff_sim} and Section \ref{sec:lens_design}). This measured efficiency depends on our assumption of the pinhole aperture efficiency, which was estimated to be $\eta_\mathrm{ap} = 0.38$ in Section~\ref{sec:experimental_setup}, but which has not yet been measured. A moderate error in this estimate could noticeably impact the inferred optical efficiency. For example, a $20\%$ reduction in the assumed $\eta_\mathrm{ap}$ from $0.38$ to $0.30$ would increase the inferred optical efficiency to $\eta_\mathrm{opt} = 40\%$, bringing it into agreement with the simulations. Future work will involve bringing the blackbody much closer to the pinhole aperture, thereby increasing $\eta_\mathrm{ap}$ to above $90\%$, and reducing the impact of any errors in this number on the optical efficiency estimate.

The discrepancy between the measured and simulated absorption efficiency may also indicate a reduced efficiency of the lens-coupled absorber. For example, the $\Pi$-shaped absorber described in Section \ref{sec:absorber_design} was optimized assuming a sheet resistance of $R_s=1.15$ $\Omega/\square$. If the aluminum film deposited here has a lower sheet impedance the absorber will couple more poorly to the incident wave. Profile errors in the silicon lens may also contribute to a reduced efficiency. The surface profile measurements described in Section \ref{sec:lens_design} constrained the radius of curvature of the lens, but did not measure the overall thickness of the lens wafer. An error in this thickness would result in a defocusing of the concentrated light. Additionally, the surface roughness was extracted from a 1-D scan of the lens array, which may not have captured the full roughness pattern. Future work will be required to measure the low temperature sheet impedance of the aluminum film, and to better describe the surface profile of the silicon lens. We note that regardless of the source of this diminshed coupling, the NEP shown in Figures \ref{fig:NEPA} and \ref{fig:nep_spectrum} is measured with respect to absorbed power, and is thus a measure of the intrinsic detector sensitivity.

We may also ask to what extent the small inferred optical efficiency may be a consequence of an error in the above analysis. The comparison of the measured $\mathrm{NEP_\mathrm{inc}}$ with the photon-limited $\mathrm{NEP_\gamma}$ is valid when the system is photon noise limited, and if this is not the case the extracted $\eta_\mathrm{opt}$ will be incorrect. However, there are a number of indications that the detector is indeed photon noise limited above $P_\mathrm{abs} \approx 10$ aW. These include the white shape of the noise PSD (Figure \ref{fig:optical_noise}), which argues against significant contribution from TLS or other noise sources with a $1/f$ profile, as well as the large difference between $S_\mathrm{xx}$ and $S_\mathrm{yy}$, which argues against a significant contribution from amplifier noise or any other additive term that may be expected to contribute equally to the noise in both quadratures. We also note that the measured NEP scales as $P_\mathrm{abs}^{1/2}$ (Figure \ref{fig:NEPA}), as expected for the photon-limited regime (Equation \ref{eq:nep_gamma}). As such, we consider it much more likely that the detector is indeed simply coupling to the cryogenic blackbody less strongly than expected.

\section{KID Sensitivity Model and Dynamic Range}

The measurements obtained in Section \ref{sec:measurements} characterize the sensitivity of this detector at $T=125$ mK and at optical loads up to 300 aW. In this section we use these measurements in a detector model to extrapolate the performance of this detector to higher temperatures and optical loading. 

\subsection{KID Model}

We use the KID sensitivity model previously developed in Ref. \citenum{HaileyDunsheath2021}, with parameters updated to match the detector measured here. The responsivity model is given by Equations \ref{eq:resp_exp} and \ref{eq:AB_def}. In Section \ref{sec:optical_measurements} we obtained estimates of $R_0 = 2.7 \times 10^{10}$ W$^{-1}$ and $P_0 = 101$ aW at $T = 125$ mK, where $n_\mathrm{th} << n^*$, and the $(1 + n_\mathrm{th}/n^*)$ term may be set to 1. These estimates also assumed a constant value of $S_2(\omega) = 3.4$. Using $\eta_\mathrm{opt} = 0.32$ we extend these results to adopt a responsivity model with finite $n_\mathrm{th}$ and variable $S_2(\omega)$:
\begin{subequations} \label{eq:resp_model}
\begin{align} 
\frac{dx}{dP_\mathrm{abs}}
&= \frac{R_0}{\eta_\mathrm{opt}} \frac{S_2(\omega)}{3.4} \bigg[\bigg(1 + \frac{n_\mathrm{th}}{n^*}\bigg)^2 + \frac{P_\mathrm{abs}}{\eta_\mathrm{opt}P_0} \bigg]^{-1/2} \\
&= \big[2.5 \times 10^{10}\,\mathrm{W}^{-1}\big] \, S_2(\omega) \\
& ~~~~~~~~~~~~~ \bigg[\bigg(1 + \frac{n_\mathrm{th}}{n^*}\bigg)^2 + \frac{P_\mathrm{abs}}{32\,\mathrm{aW}} \bigg]^{-1/2}. \nonumber
\end{align}
\end{subequations}

\noindent The model for $\tau_\mathrm{qp}$ developed in Section \ref{sec:optical_measurements} may be parameterized with $n^* = 20$ {\micron}$^{-3}$, and we compute $n_\mathrm{th}(T)$ using $\Delta_0 = 0.245$ meV. We calculate $S_2(\omega)$ by assuming a thermal quasiparticle distribution, with an effective temperature obtained by solving $n_\mathrm{th}(T) = n_\mathrm{qp}$ \cite{HaileyDunsheath2021}.

It is informative to compare the measured value of $R_0 = 2.7 \times 10^{10}$ W$^{-1}$ with a prediction using our best estimates of the fundamental KID parameters in Equation \ref{eq:AB_def}. We assume the inductor volume is $V = 20.6$ {\micron}$^3$, as designed, and make use of the results of Section \ref{sec:dark_measurements} to specify $\alpha = 0.95$ and $\Delta_0 = 0.245$ meV. The model for $\tau_\mathrm{qp}$ developed in Section \ref{sec:optical_measurements} indicates $\tau_\mathrm{max} = 420$ $\mu$s, and we adopt a pair-breaking efficiency of $\eta_\mathrm{pb} = 0.5$ (Section \ref{sec:pairbreak}). These parameters give an expected value of $R_0 = 1.9 \times 10^{10}$ W$^{-1}$, a factor of $\approx$$1.5$ smaller than measured. Equation \ref{eq:AB_def} is the product of a number of terms, each with their own uncertainty, and we are encouraged that the responsivity computed from the basic detector parameters is close to the measured value. We stress that the responsivity used in this section is based on the measurements in Section \ref{sec:optical_measurements}, and not on a first principles detector model.

At low optical loading the electrical noise at 10 Hz is $S_\mathrm{xx} \approx 5\times10^{-17}$ Hz$^{-1}$, and is assumed to be dominated by TLS noise (Figure \ref{fig:optical_noise}). The amplifier noise contribution is $S_\mathrm{xx} \approx 3.5\times10^{-18}$ Hz$^{-1}$, which corresponds to an effective noise temperature of $T_n = 15$ K when referenced to the input of the cryogenic amplifier. We expect this will be improved in the PRIMA readout chain, but conservatively carry this as-measured noise in the analysis below. The scaling of TLS noise, amplifier noise, thermal GR noise, and microwave generation noise with temperature, optical loading, and resonator detuning are all described in Ref. \citenum{HaileyDunsheath2021}. 


\subsection{Temperature Dependence}

\begin{figure}
\centering
\includegraphics[trim = 50 60 60 60, width=\singlecolwidth]{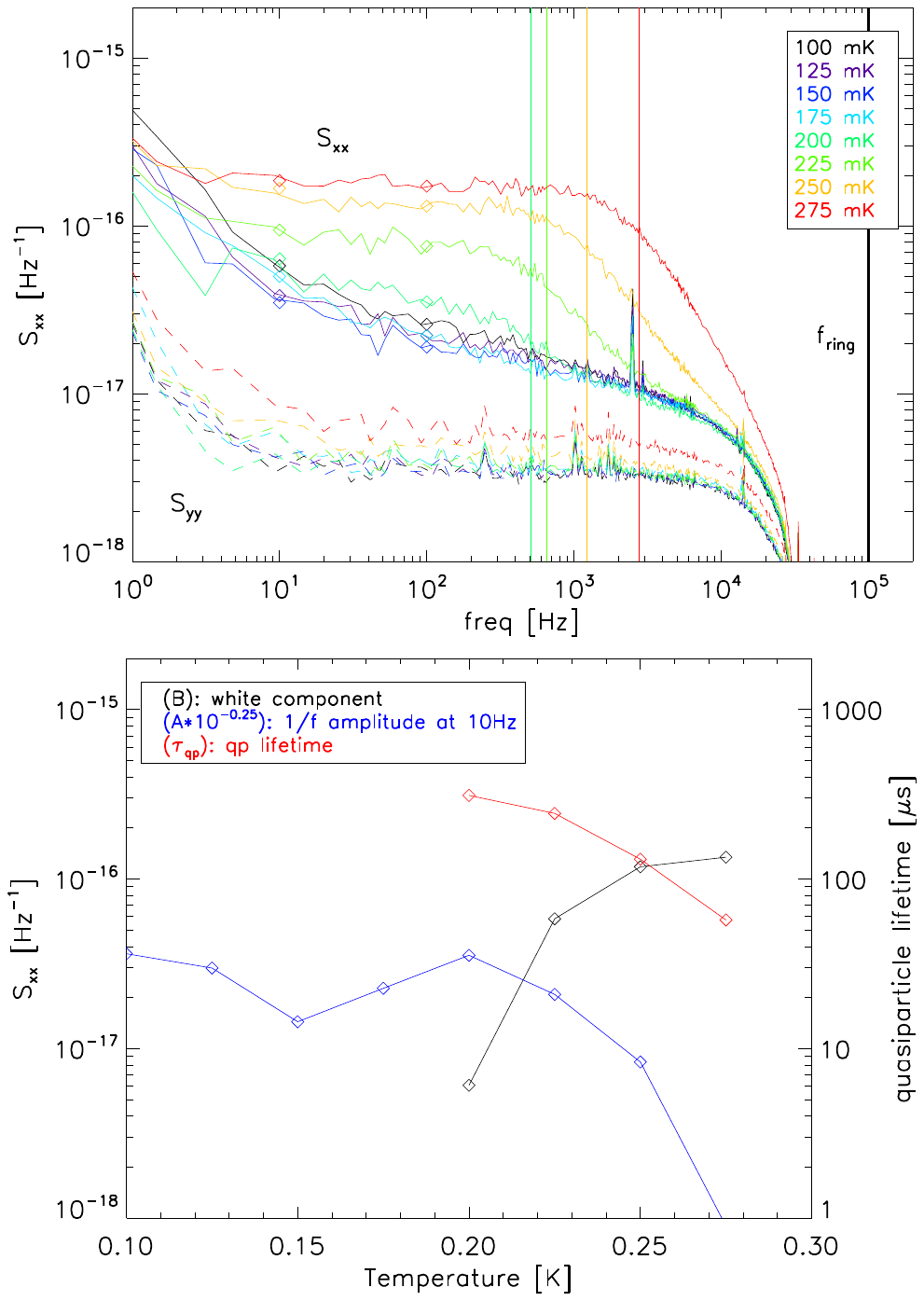} 
\caption{\small (\textit{top}) Fractional frequency noise over a range of stage temperatures, with electrical noise spikes at 60 Hz and harmonics removed. Noise in the orthogonal direction shown as dashed lines ($S_\mathrm{yy}$). Vertical lines indicate the fitted roll-off of the resonator component, and thick line at 100 kHz indicates the resonator ring-down frequency. (\textit{bottom}) Fitting results of the PSDs using Equation \ref{eq:sxx_form}.
\label{fig:dark_noise}
}
\end{figure}

\begin{figure}
\centering
\includegraphics[trim = 50 300 60 60, width=\singlecolwidth]{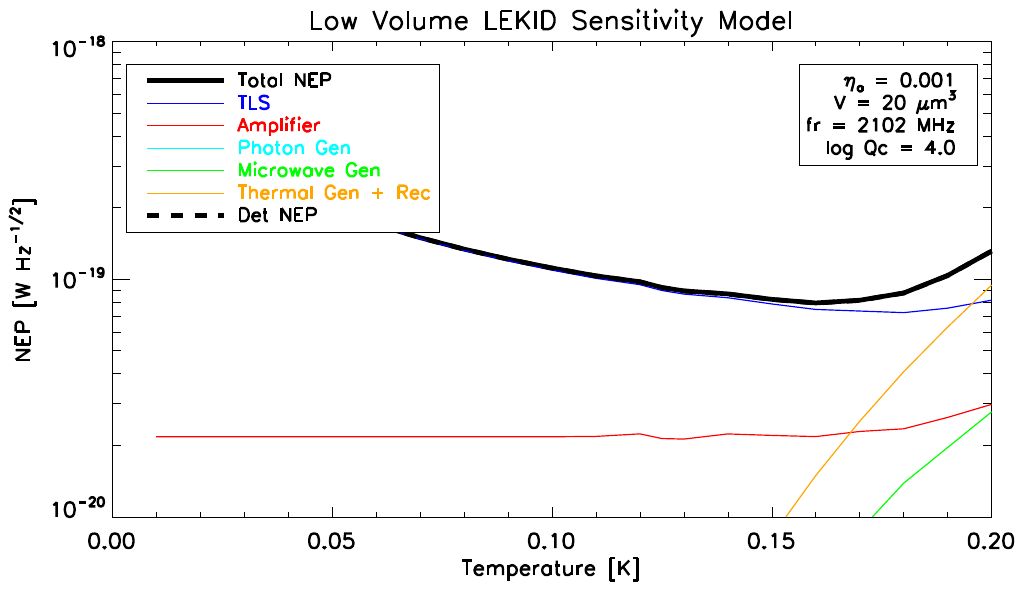} 
\caption{\small Modeled 10 Hz NEP as a function of temperature in the low optical loading limit. The TLS-limited NEP dominates at low temperature, while thermal GR noise dominates above $T \gsim 200$ mK.
\label{fig:nep_calc_temp}
}
\end{figure}

In Figure \ref{fig:dark_noise} we show the set of $S_\mathrm{xx}$ noise measurements obtained at $T=100-275$ mK for the detector in the dark condition, along with a decomposition of the PSD using the fitting function described in Equation \ref{eq:sxx_form}. At low temperature the noise at 10 Hz is dominated by the TLS component, and the rising thermal GR component is seen to dominate above $T \gsim 200$ mK. Our corresponding NEP model, which includes the reduced responsivity at higher temperatures, is shown in Figure \ref{fig:nep_calc_temp}. These figures suggest that the optimal temperature for the PRIMA detectors is about 150 mK, and that they can operate at up to $\approx$$175$ mK before starting to suffer from an increasing thermal GR noise.

\subsection{Pair-Breaking Efficiency} \label{sec:pairbreak}

Of particular interest in understanding the performance of these KIDs at THz frequencies is estimating what fraction of the absorbed photon energy is ultimately converted to quasiparticles. This is the pair-breaking efficiency $\eta_\mathrm{pb}$, which is typically assumed to be $\eta_\mathrm{pb} \approx 0.3 - 0.6$\cite{Guruswamy2014,Guruswamy2015}, but which is not often measured.

We can estimate this efficiency by comparing the GR noise in two different regimes: i) when the quasiparticle system is in thermal equilibrium, and ii) when there is a strong excess of photo-produced quasiparticles. An examination of Figure \ref{fig:reso_shift_comparison} shows that the total frequency shift resulting from varying the stage temperature with a fixed optical load is larger than the total frequency shift obtained by varying the optical load at fixed stage temperature. At the same time, Figures \ref{fig:optical_noise} and \ref{fig:dark_noise} show that the high optical loading measurements produce the most GR noise. This may be understood by noting that the resonator frequency shift is determined by $n_\mathrm{qp}$, and that for a given quasiparticle production rate (and hence equilibrium quasiparticle number density), the absorption of high energy photons produces a larger variance in $n_\mathrm{qp}$ than does thermal fluctuations, because single photons can produce multiple quasiparticles.

Our model for the GR noise in our KIDs combines photon generation noise, thermal generation noise, and recombination (of all quasiparticles) noise \cite{zmuidzinas2012,haileydunsheath2018}:
\begin{equation}
\begin{split}
S_\mathrm{xx} = \bigg(\frac{\delta x}{\delta n_\mathrm{qp}}\bigg)^2 \bigg[ \bigg( \frac{\eta_\mathrm{pb} \tau_\mathrm{qp}}{\Delta_0 V} \bigg)^2 2h\nu P_\mathrm{abs}(1 + n_0) \\ + \frac{4(\tau_\mathrm{qp})^2}{V^2} (\Gamma_\mathrm{th} + \Gamma_r ) \bigg],
\end{split}
\end{equation}

\noindent where $\Gamma_\mathrm{th}$ is the thermal generation rate, $\Gamma_r$ is the recombination rate, and we have ignored the small contribution due to readout power dissipation. In the limit of no optical loading we have $\Gamma_\mathrm{th} = \Gamma_r$, and this becomes:
\begin{equation} \label{eq:sxx_dark}
S_\mathrm{xx} = \bigg(\frac{\delta x}{\delta n_\mathrm{qp}}\bigg)^2 \bigg[ \frac{8(\tau_\mathrm{qp})^2}{V^2} \Gamma_r \bigg].
\end{equation}


\noindent In the opposite limit, if the quasiparticle population is dominated by photon absorption with little thermal generation, the recombination rate is equal to the photon generation rate ($\Gamma_r = \eta_\mathrm{pb} P_\mathrm{abs} / \Delta_0$), and $\Gamma_r >> \Gamma_\mathrm{th}$. Neglecting thermal generation, the GR noise may then be written as:
\begin{subequations} \label{eq:sxx_highload2}
\begin{align} 
S_\mathrm{xx} &= \bigg(\frac{\delta x}{\delta n_\mathrm{qp}}\bigg)^2 \bigg[ \bigg( \frac{\eta_\mathrm{pb} \tau_\mathrm{qp}}{\Delta_0 V} \bigg)^2 2h\nu \frac{\Delta_0 \Gamma_r}{\eta_\mathrm{pb}}(1 + n_0) \\ \nonumber
&~~~~~~~~~~~~~~~~~~~~~~~~~+ \frac{4(\tau_\mathrm{qp})^2}{V^2} (\Gamma_r ) \bigg] \\ 
&= \bigg(\frac{\delta x}{\delta n_\mathrm{qp}}\bigg)^2 \bigg[ \bigg(\frac{\eta_\mathrm{pb}h\nu}{2\Delta_0} \bigg) (1 + n_0) + 1 \bigg] \frac{4(\tau_\mathrm{qp})^2}{V^2} \Gamma_r.
\end{align}
\end{subequations}

\noindent The recombination rate depends on $n_\mathrm{qp}$ and $\tau_\mathrm{qp}$ as: $\Gamma_r = (n_\mathrm{qp}V/2) (\tau_\mathrm{max}^{-1} + \tau_\mathrm{qp}^{-1})$ \cite{zmuidzinas2012}. If we compare the noise measured under an optical load (following Equation \ref{eq:sxx_highload2}) with that measured dark (following Equation \ref{eq:sxx_dark}), and take care to conserve $n_\mathrm{qp}$ and $\tau_\mathrm{qp}$, we expect a ratio:
\begin{equation} \label{eq:sxx_ratio}
R = \frac{S_\mathrm{xx,optical}}{S_\mathrm{xx,dark}} = \frac{1}{2} \bigg[ \bigg( \frac{\eta_\mathrm{pb} h\nu}{2\Delta_0} \bigg) (1+n_0) + 1\bigg].
\end{equation}

\begin{figure}
\centering
\includegraphics[trim = 50 280 50 50, width=\singlecolwidth]{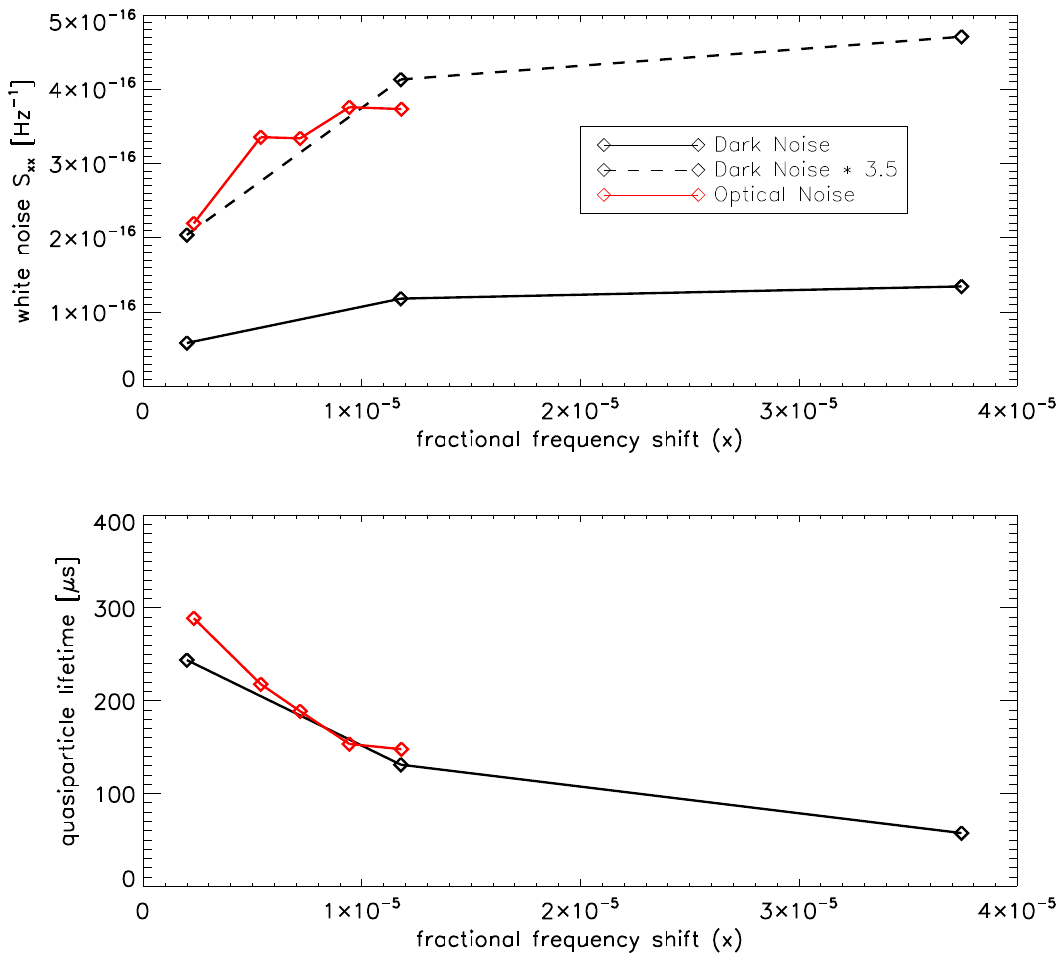} 
\caption{\small Comparison of white noise (\textit{top}) and quasiparticle lifetime (\textit{bottom}) under dark stage temperature sweeps and with elevated optical load. For a given fractional frequency shift the GR noise evident under an optical load is $\approx$$3.5$ times larger than when at elevated stage temperature, while the quasiparticle lifetime is approximately the same in the two regimes.
\label{fig:reso_shift_comparison2}
}
\end{figure}

In the top panel of Figure \ref{fig:reso_shift_comparison2} we compare the white noise level in the two regimes as a function of frequency shift. We see that for a given frequency shift the GR noise under an elevated optical load is $\approx$$3.5$ times larger than when the frequency shift is due to an increased temperature. The bottom panel compares the quasiparticle lifetime estimated in the two regimes. Here we see that $\tau_\mathrm{qp}$ is much more similar in the two cases, as expected if $\tau_\mathrm{qp}$ and the frequency shift $x$ are both primarily determined by the equilibrium $n_\mathrm{qp}$. From Equation \ref{eq:sxx_ratio}, we see that a measured ratio of $R=3.5$ corresponds to $\eta_\mathrm{pb} = 0.5$ for $\Delta_0 = 0.245$ meV, consistent with the assumption made in this work. This value is somewhat larger than the value of $\eta_\mathrm{pb} = 0.3$ that we estimate based on a calculation of the phonon escape time of 0.05 ns for our 30 nm film \cite{Kaplan1979}, and the literature value of the pair breaking time of aluminum of 0.26 ns \cite{Guruswamy2014}.

\subsection{In-Orbit Loading and Dynamic Range}

The loading for PRIMA's FIRESS spectrometer varies across the full wavelength range, as both pixel size (solid angle) and spectral bin size are subject to optimization. The instrument design prioritizes high efficiency and couples both polarizations, resulting in a per-pixel loading of $2-10$~aW when viewing low-zodiacal sightlines such as the ecliptic poles. The KIDs can be designed for higher loadings (e.g. for larger spectral bandwidth as designed for PRIMA's imager PRIMAger) simply by increasing the volume of the inductor. This will reduce the detector responsivity, but increase the maximum optical load under which the detector remains photon noise limited.

\begin{figure}
\centering
\includegraphics[trim = 50 340 50 180, width=\singlecolwidth]{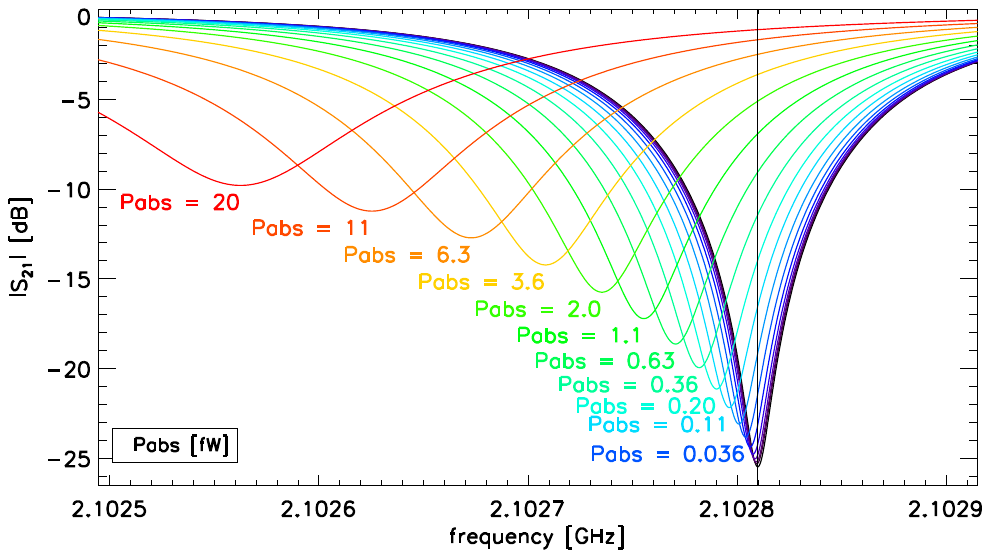} 
\caption{\small Simulated transmission profiles for this resonator under increasing optical load, up to an absorbed power of 20 fW. When operating at increased loading the detector is assumed to be biased at a fixed drive power and frequency (\textit{vertical black line}).
\label{fig:dr_fig1}
}
\end{figure}

\begin{figure}
\centering
\includegraphics[trim = 50 330 60 170, width=\singlecolwidth]{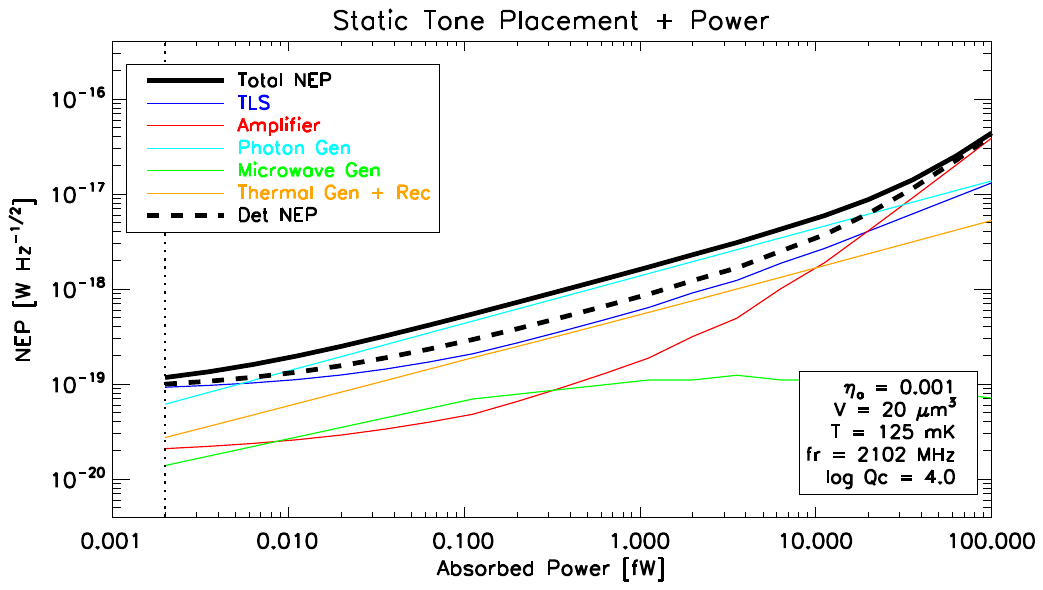} 
\caption{\small Modeled 10 Hz NEP as a function of optical loading. The detector is predicted to be photon noise limited above $\approx$$5$ aW of loading, and remain so until $\approx$$20$ fW, at which point the amplifier noise becomes dominant.  The photon noise curve applies to $\lambda=210$ \micron. At PRIMA's shorter wavelengths there is more shot noise for a given optical load, and we anticipate the short wavelength detectors may remain photon noise limited to yet higher power.
\label{fig:dr_fig2}
}
\end{figure}

To understand the capability of PRIMA KIDs to measure very bright sources we use our detector model to extrapolate the detector frequency and $Q_r$ response to higher loadings than the maximum power of 300 aW probed here. In Figure \ref{fig:dr_fig1} we show a set of modeled detector transmission profiles under increasing optical load.  
To estimate the NEP at each loading value we scale each component in the noise budget following the scalings described in Ref. \citenum{HaileyDunsheath2021}. We assume that the detector is biased with a drive power and tone frequency appropriate for operation in the low loading limit. Importantly, the noise scaling accounts for the fact that the probe tone frequency becomes progressively further detuned from the resonance frequency as the optical loading increases. The modeled noise budget is shown in Figure \ref{fig:dr_fig2}. The detector becomes photon noise limited at an absorbed power of $\approx$$5$ aW, consistent with the measurements presented in Figure \ref{fig:optical_noise}. This model predicts that the detector will remain photon noise limited until a loading of $\approx$$20$ fW. At this point the resonator becomes significantly detuned (Figure \ref{fig:dr_fig1}), and a given fractional frequency shift will have a decreasing impact on the microwave transmission. As the amplifier noise adds to the microwave signal, this noise term begins to dominate.  For reference to astronomical sources, PRIMA's FIRESS spectrometer at $\lambda$~=~100~\micron\ has a conversion of 10~milli-Janskys per aW, so the 20~fW power at which amplifier noise becomes dominant corresponds to 200~Jy.   

Even larger signals can be accommodated by the PRIMA KIDs. One approach is to simply incur the amplifier noise penalty shown in Figure~\ref{fig:dr_fig2}. Alternatively, to recover shot noise limited performance, the readout frequencies can be retuned to more closely match the loaded condition (see Figure~\ref{fig:dr_fig1}) -- this is envisioned for PRIMA when viewing bright sources. The ultimate limitation for KIDs when retuning is employed is the reduced resonator quality factor and the associated resonator crosstalk, which slowly degrades yield. These are aspects which are under consideration in our array optimization.


\section{Summary}

We describe the design, fabrication, and measurement of a far-IR kinetic inductance detector optimized to work at $210$ {\micron}. At low optical loading this detector achieves a noise equivalent power of $9\times10^{-20}$ \nepunits, meeting the sensitivity target for moderate resolution spectroscopy with the proposed PRIMA space mission. The measurements presented here suggest this detector will remain photon noise limited in the $0.005 - 20$ fW loading range, offering the capability of measuring bright astronomical sources. As companion papers describe \cite{foote_ltd2023,kane_ltd2023,cothard_ltd2024}, we are proceeding to fabricate kilo-pixel arrays based on this detector design, with initial results indicating high yield and success in bonding with lens arrays.

\section*{Acknowledgments}

The authors would like to thank Jonas Zmuidzinas for contributing to useful discussions and providing constructive criticism of the manuscript.





\end{document}